\documentclass[preprint,amscd,amsmath,amssymb,verbatim,showpacs]{revtex4}

\usepackage{graphicx}

\begin{document}

\title{Gravitatomagnetic Analogs of Electric Transformers}

\author{John Swain}
\affiliation{Department of Physics, Northeastern University, Boston, MA 02115, USA}
\email{john.swain@cern.ch}
\date{June 29, 2010}

%%%%%%%%%%%%%%%%%%%%%%%%%%%%%%%%%%%%%%%%%%%%%%%%%%%%%%%%%%%%%

\begin{abstract}
Linearized general relativity admits a formulation in terms of gravitoelectric
and gravitomagnetic fields that closely parallels the description of the electromagnetic
field by Maxwell's equations. For steady mass currents, this formalism has been
used to understand gravitomagnetic effects like the Lense-Thirring dragging of inertial
frames. For time-varying mass-energy currents, the analog of Faraday's law suggests new
effects based on the gravitational equivalent of a transformer where such currents take
the place of electrical currents. New experimental possibilities are suggested including
a novel coupling mechanism of electromagnetism to gravity, new tests of general relativity
in the ultrarelativistic limit using particle beams in the LHC, and searches for a materials
exhibiting the gravitational analog of ferromagnetism.
\end{abstract}

\maketitle

\clearpage

%%%%%%%%%%%%%%%%%%%%%%%%%%%%%%%%%%%%%%%%%%%%%%%%%%%%%%%%%%%%%%%%%%%%%%%%%%%%
%\twocolumn
\section{Introduction}

General relativity can be treated in the weak-field (linearized) limit
via a set of equations formally analogous to Maxwell's equations for electromagnetism. 
These are derived (following the conventions of \cite{Bini-et-al}; see also \cite{Mashhoon} ) 
by expanding $g_{\mu\nu}=\eta_{\mu\nu}+h_{\mu\nu}$ where $\eta_{\mu\nu}$
is the flat Minkowski metric and $h_{\mu\nu}$ represents first order perturbations.
Defining  $\bar{h}_{\mu\nu}= h_{\mu\nu}-\frac{1}{2}h\eta_{\mu\nu}$
with $h=trace(h_{\mu\nu})$ and indices raised and lowered with $\eta_{\mu\nu}$,
and neglecting terms of order $c^{-4}$, one finds the metric

\begin{equation}
ds^2=-c^2 \left( 1-2\frac{\Phi}{c^2} \right) dt^2 - \frac{4}{c}\left( \vec{A}\cdot\vec{dx} \right) dt + \left( 1+ 2\frac{\Phi}{c^2} \right)\delta_{ij}dx^i dx^j
\end{equation}

The quantities $\Phi$ and $\vec{A}$ are the gravitoelectromagnetic analogs of the scalar
and vector potential in electrodynamics.   One can define an electrogravitic field $\vec{E}$
and gravitomagnetic field $\vec{B}$ by $\vec{E}=-\vec{\nabla}\Phi-\frac{1}{c}\frac{\partial}{\partial t}(\frac{1}{2}\vec{A})$
and $\vec{B}=\vec{\nabla}\times \vec{A}$. With the gauge
condition \hbox{$\frac{1}{c}\frac{\partial \Phi}{\partial t} + \vec{\nabla}\cdot \left( \frac{1}{2} \vec{A}\right)=0$} which 
is connected to the conservation of $T^{\mu\nu}$, $\vec{E}$ and $\vec{B}$  satisfy the Maxwell-like equations:

\begin{equation}
\label{eq:faraday}
\vec{\nabla}\times\vec{E} =  -\frac{1}{c}\frac{\partial}{\partial t} \left( \frac{1}{2}\vec{B} \right), 
 \hspace*{4cm} \vec{\nabla}\cdot\left(\frac{1}{2}\vec{B}\right)=0
 \end{equation}

\noindent and

\begin{equation}
\label{eq:forflux}
\vec{\nabla}\times\left( \frac{1}{2} \vec{B} \right) = \frac{1}{c}\frac{\partial\vec{E}}{\partial t} + \frac{4\pi G}{c}\vec{j}, 
\hspace*{3cm} \vec{\nabla}\cdot\vec{E}=4\pi G \rho
\end{equation}

\noindent where $T^{00}=\rho c^2$,\, $T^{0i}=c j^i$, and $j^\mu = (c\rho,\vec{j})$ is the mass-energy current of the source,
$T^{\mu\nu}$ is the stress-energy-momentum tensor, and $G$ is Newton's constant.
Within the approximations used, 
this allows one to transcribe results from classical electrodynamics to general relativity provided
one defines sources of mass $M$ to have gravitoelectric charge $Q_E=GM$ and
gravitomagnetic charge $Q_B=2GM$. Test particles are assigned charges of the
opposite sign to ensure that gravity is attractive.

Most calculations of gravitomagnetic near-field ({\em i.e.} excluding 
gravitational radiation) effects in linearized gravity have assumed steady
mass currents. For example, the Lense-Thirring  \cite{Lense-Thirring}
dragging of inertial frames can be thought of as the effect of a steady gravitomagnetic
field (for a simple derivation, see \cite{LT-deriv}).

Experimental consequences of Faraday-law (the first expression in equation \ref{eq:faraday}) effects 
have been considered in perturbations of orbits \cite{orbits-pert} and
effects of a massive rotating object rotating moving
near a torsional oscillator\cite{Braginsky}. 

Remarkably, however, the simple transcription of an electrical transformer\cite{transformers} into
a gravitational one does not seem to have been mentioned explicitly in the literature, so 
I claim:

{\em There is a gravitational analog of an electrical transformer with 
the time-varying electrical currents through wire windings replaced by mass-energy
currents through suitable conduits. Such a transformer can be used to step up or step
down the ``gravitomotive force'' ${\mathcal{G}}$ (the line integral of $\vec{E}$ defined above
by analogy with electromotive force).}

Bini {\em et al.} \cite{Bini-et-al} consider the gravitational Faraday effect for two connected,
concentric tubes of fluid encircling a rotating mass whose angular momentum
changes linearly in time, but there is no genuine primary winding.
Forward \cite{Forward1} considers
gravitomagnetic fields produced by a time-varying mass current in order to generate
dipole ``antigravity'' type fields, but there is no
secondary winding.

Note that the operation of such a transformer is quite independent of the value of
Newton's constant, just as the operation of a normal transformer is independent
of the coupling strength of the electromagnetic field to charge.

Such a device may be difficult to construct in the laboratory, but there is an
astrophysical analog if there are flows of discrete masses. Note the difference
between a steady current (analogous to a steadily rotating mass) and the flow
of discrete masses in orbit around a central mass -- the effect described here would
be missed by approximating a flow of discrete chunks of mass  by the flow of a continuous
fluid, much as a transformer will 
work with a pulsed DC current (which clearly has
oscillating Fourier AC components) but not with a continuous DC current. 
In particular, one can imagine the gravitational analog of betatron-type
acceleration resulting from changing gravitomagnetic fluxes due to astrophysical
processes involving the motion of large masses.
Unlike gravitational radiation, the effect here is a near-field
one which avoids suppressions by powers of $v/c$ (where $v$ is a typical
speed and $c$ the speed of light) due to retardation
effects and does not require rapidly changing mass quadrupole moments\cite{MTW}.

An even more interesting possibility is that: 

{\em One can replace the primary of such a device
with a mass flow produced by non-gravitational means, such as an electrically
driven current of charged particles. In this case the output
gravitomotive force ${\mathcal{G}}$
per turn of secondary winding would be (following \cite{Mashhoon})
\begin{equation}
{\mathcal{G}}= -\frac{2}{c}\frac{d}{dt} \int \vec{B}\cdot \vec{da}
\end{equation}
where the integral on the right hand side is the gravomagnetic flux produced by
the source from the first equation in \ref{eq:forflux} that links the secondary.
For example, a time-varying flow of charged massive particles
could be driven through the primary by electromagnetic
forces. This would give rise not only to the usual Maxwell magnetic fields, but also to 
gravitomagnetic fields.
Effectively one has a transformer which converts a time-varying electromotive force
into a time-varying gravitomotive force which can, in principle, be
made arbitrarily large by increasing the number of turns on the secondary. }

Just as is the case for electrical transfomers\cite{transformers},
one can imagine tuning the primary and/or secondary to (electrical or
mechanical) resonance with the driving force. This technique is used commonly
for efficient high-frequency air-core transformers including Tesla coils\cite{Tesla} and
would be natural for for situations where there is no gravitational analog
of a ferromagnetic or even ferrite core. Any electromagnetic Faraday-type couplings between primary and
secondary can be prevented by magnetically shielding them from each other allowing the
gravitomagnetic inductive effects to be clearly separated. In addition, the flows in the secondary
need not be of electrically charged particles, again eliminating any effects due to conventional
electromagnetic induction. Clearly the situation with primary and secondary exchanging
roles can also be envisaged, although it is hard to envision significant laboratory sources of
gravitationally driven electrically neutral matter.

Extremely relativistic cases such as a pulsed beam of (laser) light
or the pulsed beam of particles in an accelerator as the primary are particularly
interesting. Approximations for the Maxwell-like description
neglecting powers of $v/c$ may not be valid \cite{notvalid}
and new and distinct effects might be
found experimentally before being predicted theoretically. Braginsky {\em et al.}
\cite{Braginsky} consider experiments to measure the gravitational kick
delivered by a relativistic particle bunch as it passes a suitable detector, but
this is distinct from the induction effect suggested here. 

Laboratory experiments to test relativistic
gravitational effects\cite{Braginsky} are typically hampered by the difficulty of making large
masses move at large speeds. For beams in an accelerator such as the LHC \cite{LHC}
one has particles moving very close to the speed of light so all effects usually
suppressed by powers of $v/c$ (including the leading suppression of magnetic
over electric effects by one power of $v/c$) are now unsuppressed. In addition, very high accelerations
(very high angular velocities in the case of circular motion) and thus high
rates of change of gravitomagnetic flux are possible which could not be achieved
with normal matter: the up to 7 TeV protons in the
LHC are $10^{11}$ protons per bunch times 2808 bunches with a
relativistic $\gamma=\frac{1}{\sqrt{1-v^2/c^2}}\approx 7000$
revolving in a 27 km circle at a rate of 40 MHz, held together in a way
that no ordinary matter could via external electromagnetic fields. 

The LHC is cooled by superfluid liquid
helium at 1.9 K, so much of the infrastructure for
superconducting magnetic shielding or
cryogenic detectors - in particular superconducting\cite{supercJosephson} or
superfluid\cite{superfJosephson} Josephson
interferometers - is already present. Detailed proposals for various
experimental scenarios are in preparation\cite{me}.

The production
of high frequency gravitational waves by accelerators 
such as the LHC\cite{LHC} or the Tevatron\cite{Tevatron-GW} has
been considered previously \cite{Palazzi}. Taking the source
of magnetogravitic flux to be a gravitational wave produced by an accelerator,
the secondary of the proposed transformer would essentially be a
loop antenna, but it should be noted that what is proposed here 
a new effect which is strictly near-field and distinct from gravitational radiation.
Gravitational sychrotron radiation produced would
necessarily be accompanied by much larger electromagnetic synchrotron
radiation losses in any readily conceivable device so that fact that the
effect considered here is near-field (not radiation) is very important.

Eddy current losses due to induced EMF's can, of course, be
made very small independent of any near-field gravitational effects
via standard techniques\cite{transformers}.

As noted long ago by Forward \cite{Forward1,Forward2},  electrical
transformers are remarkably efficient at transferring power.  A large part of this efficiency is due to 
the existence of ferromagnetic materials with very high magnetic permeability
which can be used to make essentially all the magnetic flux produced by
the primary link the secondary. He suggested that a search be made for
the gravitational analog of such materials, pointing out that ferromagnetic
levels of permeability are ultimately due to cooperative effects of spins.
Whether or not spin coupling to torsion\cite{torsionreview}
might play a role could be open here to experimental tests. Torsion is zero
in standard general relativity.
Certainly gravitational paramagnetic and diamagnetic effects have
been considered in the literature\cite{paradia,alyogi}. Thus:

{\em One can use a transformer of the kind described above (for example
with pulsed electric current or particle beam in the primary and a superfluid
or superconducting secondary) to measure, or at least set bounds on, the
gravitational equivalent of magnetic permeability, and to search for the
gravitational equivalent of ferromagnetic materials by using them as cores.}

So far\cite{exp} attempts have only been made using
wedges of material placed between laboratory sources of
gravitational waves and Weber-type\cite{Weber} detectors.

One should keep in mind that ferromagnetism is a very unusual state of condensed matter
and its  early discovery was very fortuitous, relying 
on the fact that some materials exhibiting it were easy to find in nature. 
Without this good luck, magnetism itself might have been very hard to discover. 

Other surprising and unforeseen phenomena in bulk condensed matter systems such
as superconductivity\cite{superconductivity-disc} and superfluidity\cite{superfluidity-disc} were
found much later, with
colossal magnetoresistance only discovered in the 1950's\cite{CMR},
and Bose-Einstein condensation in 1995\cite{BE}. 

Aside from analogs of
ferromagnetism, other effects such as effective $\vec{E}\cdot\vec{B}$ couplings
in condensed matter might also be looked for. However unlikely such effects
might seem at the moment, there are simply no constraints on their existence
at all at present.
Given that there are no predictions even
of the magnetic permeability of ferromagnetic materials from first principles, 
one might be optimistic that an as-yet-undiscovered gravitational
analog might be found - only experiment can tell!

\section{Acknowledgements}

I would like to thank Ka\'{c}a Bradonji\'{c} and Tom Paul for reading an early
draft of this paper. This work
was supported in part by the US National Science Foundation
and grant NSF0855388.


\begin{thebibliography}{99}
%-----------------------------------------------------------------------------
\bibitem{Bini-et-al} D. Bini, C. Cherubini, C. Chicone, and B. Mashhoon,
Class. Quant. Grav. {\bf 28} (2008) 225014.
\bibitem{Mashhoon} B. Mashhoon, ``Gravitoelectromagnetism'', in {\em References Frames and Gravitomagnetism},
ed. J.-F. Pascual-S\'{a}nchez {\em et al.}, World Scientific (Singapore) p121; B. Mashoon, ``Gravitomagnetism: a brief review'',
in {\em The Measurement of Gravitomagnetism: A Challenging Enterprise}, ed. L. Iorio, Nova Science (New York), 2007.
\bibitem{transformers} Standard references include R. Lee {\em et al.}, {\em Electronic Transformers and Circuits, 3$^{rd}$ Edition},
Wiley-Interscience, 1988 and  Colonel William T. McLyman, {\em Transformer and Inductor Design Handbook}, CRC Press, 2004;
\bibitem{Forward1} R. L. Forward, American Journal of Physics, {\bf 31} (1963) 166.
\bibitem{Forward2} R. L. Forward, Proc. IRE {\bf 49} (1961) 892.
\bibitem{Lense-Thirring} J. Lense and H. Thirring, Phys. Z. {\bf 29} (1918) 156.
\bibitem{LT-deriv} O. I. Chashchina, L. Iorio, and Z. K. Silagadze, Acta Phys. Polon. {\bf 40} (2009) 2363.
\bibitem{orbits-pert} Matteo Luca Ruggiero, Lorenzo Iorio, arXiv:0906.1281 (2009).
\bibitem{Braginsky} V. B. Braginsky, C. M. Caves and K. Thorne, Phys. Rev. {\bf D15} (1977) 2047.
\bibitem{MTW} See, for example, the very clear discussions in C. W. Misner. K. S. Thorne, J. A. Wheeler, {\em Gravitation},
W. H. Freeman; 2nd Printing edition (1973).
\bibitem{Tesla} See, for example, M. Tilbury, {\em The Ultimate Tesla Coil Design and Construction Guide}, McGraw-Hill/TAB Electronics, 2007.
\bibitem{notvalid} For example, Christodoulou has pointed out a nonlinear effect of gravitational waves on laser
interferometer detectors which results in a permanent displacement of the detector after a gravitational wave has
passed. It is of the same order of magnitude as the linear effect. See D. Christodoulou, Phys. Rev. Lett. {\bf 67} (1991) 1486.
\bibitem{LHC} See the LHC website at {\tt http://lhc.web.cern.ch/lhc/ }
\bibitem{supercJosephson} G. Papini, Phys. Lett. {\bf 24A} (1967) 32.
\bibitem{superfJosephson} J. Anandan, Phys. Rev. Lett. {\bf 47} (1981) 463;
J. Anandan and R. Chiao, General Relativity and Gravitation, {\bf 14} No. 6 (1982) 515;  J. Anandan, J. Phys. A: Math. Gen. {\bf 17} (1984) 1367.
\bibitem{me} J. Swain, in preparation.
\bibitem{Palazzi} G. Diambrini Palazzi, Phys. Lett. {\bf B197} (1987) (302); G. Diambrini Palazzi, BOLL. SOC. ITAL. FIS., NUOVA SER. 
4 {\bf 4} (1988) 51; P. Chen {\em et al.} in Los Angeles 1991, ``Remarks on the production of gravitational waves by EM radiation and particle beams'', Proceedings, Beam-beam and beam-radiation interactions'';
F. Chiarello {\em et al.}, ``On production of gravitational radiation by particle accelerators and by high power lasers'', in
Proceedings of the Edoardo Amaldi Meeting on Gravitational Wave Experiments, Rome, Italy, 14-17 Jun 1994, 282;
G. Diambrini Palazzi, Part. Accel. {\bf 33} (1990) 195; P. Chen, Mod. Phys. Lett. {\bf A6} (1991) 1069; V. I. Pustovoit and M. E. Gertsenshtein,
Sov. Phys. JETP {\bf 15} (1962) 116; O. P. Shushkov and I. B. Khriplovich, Sov. Phys. JETP {\bf 39} (1974) 1
\bibitem{Tevatron-GW} P. Reiner {\em et al.}, ``Search for Anomalous Gravitational Effects at the Fermilab Accelerator: Progress Report on E-723'',
FERMILAB-PUB-84-097-E, UR-898, Oct 1984.
\bibitem{torsionreview} For a review on torsion, which is absent in standard general relativity, see for example
F. W. Hehl {\em et al.}, Rev. Mod. Phys. {\bf 48} (1976) 393.
\bibitem{paradia} M. Agop {\em et al.}, Aust. J. Phys. {\bf 49} (1969) 1063.
\bibitem{alyogi} Y. Srivastava and A. Widom, Phys. Lett. {\bf B280} (1992) 52.
\bibitem{exp} V. B. Braginsky {\em et al.} Zhur. Eksp. i Theoret. Fiz. {\bf 43} (1962) 51.
%Sov. Phys. JETP  {\bf 16} (1963) 36.
\bibitem{Weber} J. Weber, Phys. Rev. {\bf 117} (1960) 306.
\bibitem{superconductivity-disc} H. Kamerlingh Onnes, H., Comm. Phys. Lab. Univ. Leiden (1911) Nos. 122 and 124.
\bibitem{superfluidity-disc} J. F. Allen and A. D. Misener, Nature {\bf 141} (1938) 75.
\bibitem{CMR} G. H. Jonker and J. H. Van Santen, Physica {\bf 16} (1950) 377.
\bibitem{BE}  M.H. Anderson {\em et al.} Science {\bf 269} (1995) 198.
\end{thebibliography}
\end{document}